\documentclass{article}
\usepackage{spconf,amsmath,graphicx}
\usepackage{bm}
\usepackage{subfigure}
\usepackage{float}

\title{DEEP GENERATIVE FACTORIZATION FOR SPEECH SIGNAL}
\name{Haoran Sun, Lantian Li, Yunqi Cai, Yang Zhang, Thomas Fang Zheng, Dong Wang
}
\address{Center for Speech and Language Technologies, Tsinghua University}
\begin{document}
%
\maketitle
\begin{abstract}

Various information factors are blended in speech signals, which forms the primary difficulty for most speech information processing tasks.
An intuitive idea is to factorize speech signal into individual information factors (e.g., phonetic content and speaker trait),
though it turns out to be highly challenging.
This paper presents a speech factorization approach based on a novel factorial discriminative normalization flow model (factorial DNF).
Experiments conducted on a two-factor case that involves phonetic content and speaker trait
demonstrates that the proposed factorial DNF has powerful capability to factorize speech signals and outperforms several comparative models
in terms of information representation and manipulation.

\end{abstract}
\section{Introduction}
\label{sec:intro}

Speech signal is highly complex and involves variations from multiple sources~\cite{flanagan2013speech}.
Decomposing speech signal into elementary components is perhaps the most important idea in the history of
speech processing research.

Fourier transform is such a decomposition, which decomposes
speech signals into a set of simple periodical functions, each with a single frequency.
This Fourier decomposition, also known as spectrum analysis~\cite{fulop2011speech},
reveals the frequency-dependent energy distribution,
a fundamental property behind the complex vibration of the waveform. This decomposition, however,
does not employ any knowledge on the speciality of speech signals, and so cannot offer deep understanding for speech.
The source-filter model
solves this problem by formulating speech production as an excitation and modulation process~\cite{fant1970acoustic}.
Inversely, speech signal can be decomposed into excitation and modulation according to the production model.
This excitation-modulation decomposition `was the basis of practically all the work on speech signal
processing that followed'~\cite{benesty2008introduction}.

In spite of the predominant importance, the excitation-modulation decomposition
plays a role on low-level signals and thus, is not directly related to high-level information
such as phonetic contents and speaker traits. In fact, both the excitation and modulation components
derived from the excitation-modulation decomposition are irrelated to speaker traits, which means that
this decomposition is less useful for identifying speaker identity. For the sake of speech information
processing, we hope to decompose speech signals into high-level information factors.

In this paper, we will present such a speech information decomposition approach.
We will formally define this decomposition as a nonlinear extension of factorization analysis, and
propose an implementation based on the normalization flow model.

\section{Deep generative factorization}

Fujisaki~\cite{fujisaki1997prosody} is perhaps the first scholar mentioning the
information-based production/decomposition.
In his proposal, speech signal can be regarded as a composition of three types of
information factors: linguistic factors that determine what to speak, paralinguistic factors
that determine how to speak, and non-linguistic factors that determine the residual properties that are
unrelated to the speech content.


Although the concept is clear, it is not easy to turn Fujisaki's proposal to a computational model, due to the
complex and unknown convolution amongst information factors. In fact, the concept of \emph{information factor}
per se is not easy to define. Fortunately, the recent development on deep generative models offers a renewed
possibility. Basically, we
shall represent each elementary variation $c$ (e.g., phonetic content and speaker trait)
in speech signal as \emph{a set of random variables}, denoted by $v_c$. We will call $v_c$ the information factor
corresponding to variation $c$.
By this definition, the distribution of speech signal can be modeled by a \emph{deep composition} of the
distributions among the information factors. Formally, it will take the following form:

\begin{equation}
\label{eq:fact}
x = G(v_{c_1}, v_{c_2}, ...),
\end{equation}
\noindent where $G$ is a deep generative model, and each information factor $v_{c_i}$ is a standard multivariate Gaussian.
Note that different information factors are independent. As a simplest case, we assume speech signal can be decomposed
into a phonetic content factor $v_q$, a speaker factor $v_s$, and a Gaussian noise $\epsilon$, following the linear form shown below:

\begin{equation}
\label{eq:fact:linear}
x = M_q v_q + M_s v_s + D \epsilon.
\end{equation}
\noindent where $D$ is a diagonal matrix. This can be reformulated to the following simple form:

\begin{equation}
x = \begin{bmatrix}
M_q \ M_s
\end{bmatrix}
\begin{bmatrix}
v_q \\
v_s
\end{bmatrix}
+ D \epsilon,
\end{equation}
\noindent Note that this is just the standard factorization analysis~\cite{bishop2006continuous}, where $v_q$ and $v_s$
form the factors. Therefore, the deep generation model in Eq.(\ref{eq:fact}) can be regarded as a nonlinear extension of factorization analysis.
For this reason, the decomposition according to deep generative model in Eq.(\ref{eq:fact}) can be called \textbf{deep generative factorization}.

\section{Deep factorization based on DNF}
\label{sec:method}

\subsection{Revisit NF and DNF}

Normalization flow (NF) \cite{dinh2014nice,dinh2016density,kingma2018glow} is
a popular deep generative model. The basic idea of NF is to transform a normal distribution $p(\bm{z})$ to match the data by an \emph{invertible} function $f$.
Due to the invertibility, the latent variable $\bm{z}$ can be obtained exactly by $\bm{z}=f^{-1}(\bm{x})$.
This means that an extra inference network is not necessary, and we can compute the likelihood function $p(\bm{x})$ exactly.

Suppose the latent variable $\bm{z}$ and the observation $\bm{x}$ are linked by an invertible transform $f_{\theta}$, then $p(\bm{z})$ and $p(\bm{x})$
hold the following relation~\cite{rudin2006real}:

\begin{equation}
	\log p(\bm{x})=	\log p(\bm{z})+\log \left| \det (\frac{{\rm d} {f_\theta}^{-1}(\bm{x})}{{\rm d}\bm{x}}) \right|,
\end{equation}
\noindent where $\det(\cdot)$ represents determinant of the Jacobian matrix of $f^{-1}$. In the above equation, the two terms on the right hand side
are often called the \emph{prior term} and the \emph{entropy term}, respectively.
The NF model is trained by maximizing the likelihood of the
training data with respect to the parameter $\theta$, and the log likelihood function can be computed as follows:

\begin{equation}
	\mathcal{L} (\theta) = \sum\limits_{i}{\log p({\bm{z}_{i}})}+\sum\limits_{i}{\log \left| \det (\frac{{\rm d}{f_\theta}^{-1}({\bm{x}_{i}})}{{\rm d}{\bm{x}_{i}}}) \right|}.
\end{equation}
\noindent This function can be maximized by any numerical optimization method, for instance stochastic gradient descend (SGD).

\begin{figure}[htb!]
	\centering\includegraphics[width=0.80\linewidth]{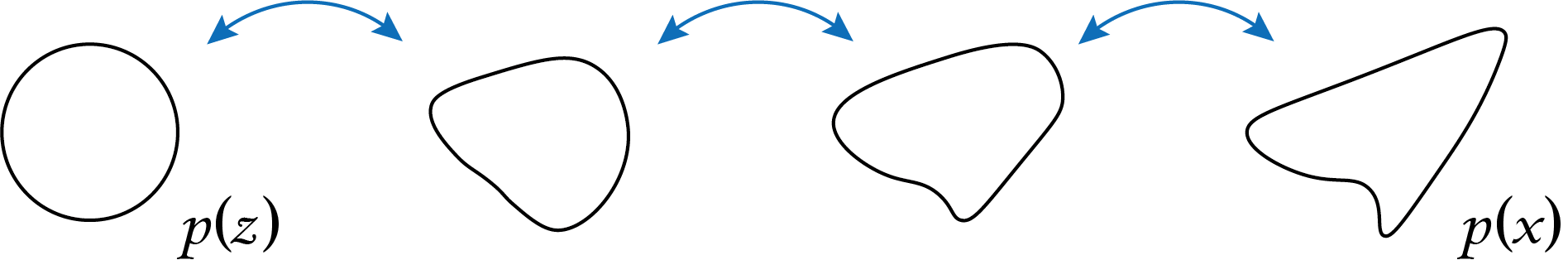}
	\caption{Distribution transform with normalization flow.}
	\label{fig:nfmapping}
\end{figure}

In practice, the invertible function $f_\theta$ is often implemented as a sequence of simple invertible functions, which transforms a
simple distribution on $\bm{z}$ \emph{gradually} to a complex distribution on $\bm{x}$, as shown in Figure~\ref{fig:nfmapping}.
It is shown that if $f$ is powerful enough, any complex distribution can be obtained from a simple Gaussian~\cite{papamakarios2019normalizing}.

\begin{figure}[htb!]
	\centering\includegraphics[width=0.80\linewidth]{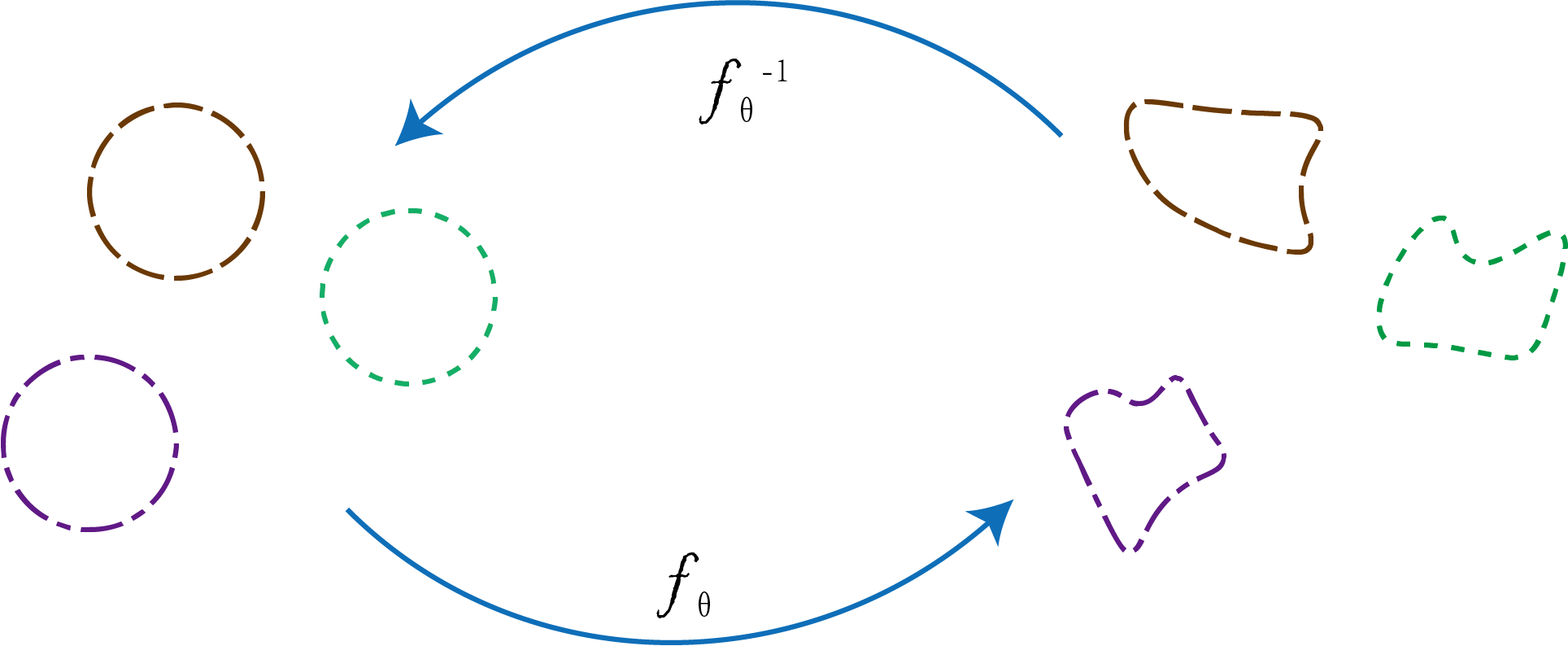}
	\caption{The DNF architecture, where each class has its individual prior distribution.}
	\label{fig:dnf}
\end{figure}

Discriminative normalization flow (DNF) \cite{cai2020deep} is an extension of the NF model.
As shown in Figure~\ref{fig:dnf}, different classes share the same NF network,
but the prior distributions are different.
Following the notations of NF,
and assuming the prior of class $y$ to be $\mathcal{N}(\bm{\mu}_y, \bm{I})$,
the probability of an observation $\bm{x}$ is:

\begin{equation}
	\log p(\bm{x})=	\log \mathcal{N}(\bm{z}; \bm{\mu}_{y(\bm{x})}, \bm{I})+ \log J(\bm{x}),
\end{equation}
\noindent where $\log J(\bm{x})$ denotes the entropy term, and $y(\bm{x})$ denotes the class label of $\bm{x}$.
Similar to NF, DNF can be trained by maximum likelihood.

\subsection{Factorial DNF}

The DNF model can encode a single information factor (corresponding to the class label).
However, our purpose is to represent multiple information factors in a full generative model.
To accommodate this request, we split the latent code into several \emph{partial codes},
with each partial code corresponding to a particular information factor.
This model is denoted by \textbf{factorial DNF}.

Taking the case of two information factors as an example, we split the latent code into two partial codes,
$\bm{z}^A$ and $\bm{z}^B$, corresponding to the information factor $A$ and $B$ respectively.
Since the prior distribution is a diagonal Gaussian, $\bm{z}^A$ and $\bm{z}^B$ are naturally independent,
which means:

\begin{equation}
p(\bm{z}) = p(\bm{z}^A)  p(\bm{z}^B).
\end{equation}

\noindent We assume that the prior distributions for $\bm{z}^A$ and $\bm{z}^B$ depend on the labels
corresponding to the information factors $A$ and $B$, respectively. More precisely,

\begin{align}
	p({\bm{z}^{A}})=\mathcal{N}({\bm{z}^{A}}; \bm{\mu }_{y_A(\bm{z})}, \bm{I})\nonumber \\
	p({\bm{z}^{B}})=\mathcal{N}({\bm{z}^{B}}; \bm{\mu }_{y_B(\bm{z})}, \bm{I}) \nonumber
\end{align}
where $y_A(\bm{z})$ and ${y_B(\bm{z})}$ are the class labels of $\bm{z}$ for factor $A$ and $B$ respectively.
The likelihood $p(\bm{x})$ can therefore be written by:

\begin{equation}
	\log p(\bm{x})= \log p(\bm{z}^A) + \log p(\bm{z}^B) + \log J(\bm{x}).
\end{equation}

Once the model has been well trained, an observation $\bm{x}$ can be encoded to $\bm{z}=[\bm{z}^A \ \bm{z}^B]$ by the invertible transform $f^{-1}$,
and the partial codes $\bm{z}^A$ and $\bm{z}^B$ encode the information factors $A$ and $B$, respectively.

\subsection{Speech factorization by factorial DNF}

Factorial DNF can be used to implement the deep generative factorization shown in Eq.(\ref{eq:fact}).
We will describe the approach with a two-factor case that involves a phone factor and
a speaker factor.

Firstly, speech signals are split into short segments and all the segments are labeled by phone and speaker classes,
denoted by $Q$ and $S$ respectively. During training, treat each short segment as an observation.
Select the partial means $\bm{\mu}_q$ and $\bm{\mu}_s$ for the partial codes $\bm{z}^Q$ and $\bm{z}^S$
according to its phone class $q$ and speaker class $s$. The prior for the latent variable $\bm{z}$ is then
formed to be a Gaussian where the mean vector is $[\bm{\mu}_q \ \bm{\mu}_s]$.
Note that the likelihood function $p(\bm{x})$ can be computed, and so the model can be trained without principle difficulties.

By this setting, we have formulated the variation in speech signal into three parts:
(1) the randomness on phone class means $\bm{\mu}_q$; (2) the randomness on speaker class means $\bm{\mu}_s$;
(3) the residual randomness, reflected by the Gaussian distribution on $\bm{z}$ conditioned on
$[\bm{\mu}_q \ \bm{\mu}_s]$.

We highlight that since the NF transform is invertible, all the variation/information in the original
speech is retained in the code $\bm{z}$. However, factorial DNF conducts an interesting `variation reshaping'
that makes variation corresponding to different information factors uncorrelated and confined in their own dimensions.
This factorization holds two significant advantages: (1) The inference is as simple as an inverse transform $f^{-1}(\bm{x})$,
which is different from conventional shallow factorization models such as JFA that requires complex Bayesian inference.
(2) The code corresponding to an information factor can be obtained easily by 
selecting its individual dimensions, and changing codes corresponding to different information factors will not impact each other.


%

\section{Related work}
\label{sec:related}

Speech factorization has been studied in speaker recognition for a long time.
The famous Gaussian mixture model-Universal background model (GMM-UBM) is an early example~\cite{Reynolds00},
which firstly represents the short-term phonetic factor by a discrete random variable represented by the
Gaussian components, and then represents the speaker factor as a mean shift on each of the components.
The succeeding subspace models such as eigenvoice model~\cite{kenny2002maximum}, i-vector model~\cite{dehak2010front}
and joint factor analysis (JFA) model~\cite{kenny2007joint} follow the same principle but regulate the variation of
speakers and sessions in a low-dimensional space.
Some researches has proposed to use deep generative models to factorize speech variation.
For example, Hsu et al.~\cite{hsu2017learning} employ a sequential
VAE to discriminate phone and speaker factors. Another noticeable work is multi-style speech synthesis using
style tokes~\cite{wang2018style}.

Our approach in this paper is based on deep generative models as~\cite{hsu2017learning}, however
it uses supervised training to ensure the quality of the factorization model, and the invertibility of the model guarantees
perfect speech reconstruction.
Note that a fully supervised deep speech factorization approach was proposed by Li et al.~\cite{li2018deep},
however it is not a generative model and so cannot guarantee a perfect reconstruction.

Our work is also related to deep speech representation, in the sense that both learn latent codes to represent speech.
Contrastive prediction coding (CPC)~\cite{oord2018representation} and autoregressive prediction
coding (APC)~\cite{chung2019unsupervised} are trained to produce latent codes that predict the future samples based on the past samples.
Our work follows the theme of latent representation learning, whereas focuses on information factorization.

\section{Experimental}
\label{sec:exp}

\subsection{Data}
The TIMIT database is used in our experiments.
The original 58 phones in the TIMIT transcription are mapped to 39 phones by Kaldi toolkit~\cite{povey2011kaldi} following the TIMIT recipe,
and 38 phones (\emph{silence} excluded) are used as the phone labels.
To balance the number of classes between phones and speakers, we select 20 female and 20 male speakers, resulting in 40 speakers in total.

All the speech utterances are firstly segmented into short segments according to the TIMIT phone transcriptions by force alignment.
All the segments are trimmed to 200ms; if a segment is shorter than 200ms, we extend it to 200ms in both directions.
All the segments are labeled by phone and speaker classes.
Afterwards, every segment is converted to a $20 \times 200$ time-frequency spectrogram by FFT, where the window size
is set to 25ms and the window shift is set to 10ms. The spectrograms are reshaped to vectors and are used as input features of
deep generative models.


\begin{figure*}[htb]
	\centering
	\includegraphics[width=0.86\linewidth]{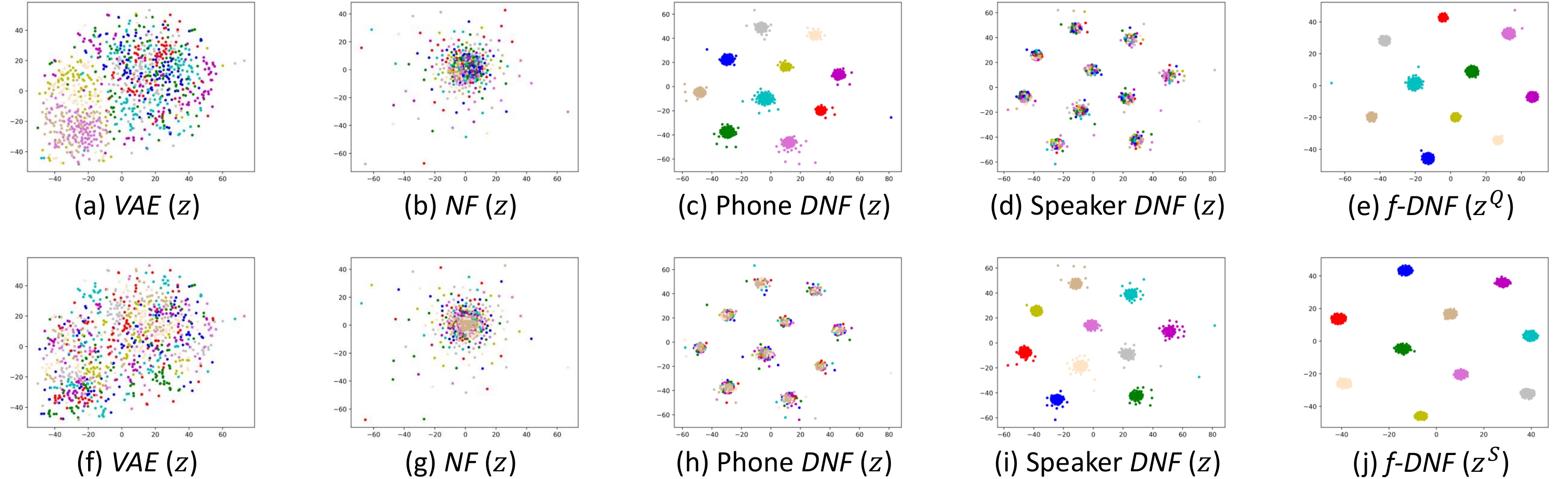}
	\vspace{-1mm}
	\caption{The latent codes generated by various models, plotted by t-SNE.
	In the first row (a) to (e), each color represents a phone; in the second row (f) to (j), each color represents a speaker.
	`Phone DNF' denotes DNF trained with phone labels; `Speaker DNF' denotes DNF trained with speaker labels.}
	\label{fig:encode}
\end{figure*}
\vspace{-1mm}

\subsection{Model settings}

The NF, DNF and factorial DNF models are based on the same \emph{RealNVP} architecture~\cite{dinh2016density}.
It involves 6 blocks, and each block contains a coupling layer and a batch norm layer.
The class means of DNF and factorial DNF are initialized from a normal distribution,
and the variance of each class is set to $I$.
For factorial DNF, the dimensions of the partial codes $\bm{z}^Q$ for phone and $\bm{z}^S$ for speaker
are equal, and both are set to 2,000. The Adam optimizer~\cite{kingma2014adam} is used to train all these models.

For comparison, we also present the results with VAE. The source code released by Hsu et al.~\cite{hsu2017learning}
is used to build the VAE model.

\subsection{Encoding}

In the first experiment, we conduct a qualitative study on the latent codes generated by VAE, NF, DNF and factorial DNF.
We use t-SNE~\cite{saaten2008} to draw the distributions of the latent codes generated by different models.
Results are shown in Figure ~\ref{fig:encode}. It can be observed that the latent codes generated by VAE and NF almost lose the class structure;
DNF can retain the class structure of the information factor corresponding to the class labels in the model training;
Factorial DNF can retain the class structure corresponding to all the information factors.


\subsection{Factor manipulation}

The second experiment tests the quality of the factorization by manipulating the factors.
Presumably, if the factorization is perfect, changing one factor will not modify the properties corresponding to other factors.

We use the mean-shift approach to conduct the manipulation.
Given a factor $A$ to manipulate, we compute the mean vectors of
each class on factor $A$, denoted by $\{\mu_{A,i}\}$.
Now for a sample $x$ from one class $c_1$, we hope to change it to another class $c_2$. This can be obtained by
moving its latent code $z$ by a shift $\mu_{A,c_2} - \mu_{A,c_1}$ and then transforming it back to the observation space.
In summary:

\vspace{-1mm}
\[
x' = f(f^{-1}(x)  + \mu_{A,c_2} - \mu_{A,c_1}).
\]

We will test the results when the phone factor and the speaker factor are manipulated respectively.
Take the phone manipulation as an example,
suppose the conversion for phone is from $q_1$ to $q_2$,
and the speaker label $s$ remains the same.
We will report the posteriors $p(q_2|x)$, $p(q_2|x')$, $p(s|x)$ and $p(s|x')$.
Experiments are conducted on all pairs of phones and the averaged posteriors are reported in Table~\ref{tab:conv}.
The same settings are applied to speaker manipulation.
We trained a speaker and a phone MLP classifiers respectively to compute corresponding posteriors.
Both classifiers contain one hidden layer with 800 hidden units.

\begin{table}[htb!]
	\caption{MLP posteriors on the target class before and after phone/speaker manipulation. `f-DNF' denotes factorial DNF.
	$\delta(\cdot)$ denotes the difference on posteriors  $p(\cdot|x')$ and $p(\cdot|x)$.}
	\label{tab:conv}
	\centering
	\scalebox{0.88}{
	\begin{tabular}{l|ccc|ccc}
    \hline
                        & \multicolumn{6}{c}{Phone Manipulation} \\
    \hline
    Model               &  $p(q_2|x)$   & $p(q_2|x')$  & $\delta(q_2)$  & $p(s|x)$  & $p(s|x')$   &  $\delta(s)$ \\
    \hline
    VAE                 & 0.013  & 0.312  & 0.299 & 0.612 & 0.454 & -0.158 \\
    NF                  & 0.013  & 0.410  & 0.397 & 0.612 & 0.489 & -0.123 \\
    DNF                 & 0.013  & 0.619  & 0.606 & 0.612 & 0.335 & -0.277 \\
    f-DNF               & 0.013  & \textbf{0.636} & \textbf{0.623} & 0.612 & \textbf{0.536} & \textbf{-0.076} \\
    \hline
    \hline
                        & \multicolumn{6}{c}{Speaker Manipulation}    \\
   \hline
    Model               &  $p(s_2|x)$   & $p(s_2|x')$  & $\delta(s_2)$  & $p(q|x)$  & $p(q|x')$  & $\delta(q)$ \\
    \hline
    VAE                 & 0.010  & 0.303 & 0.293 & 0.520 & \textbf{0.509} & \textbf{-0.011} \\
    NF                  & 0.010  & 0.435 & 0.425 & 0.520 & 0.484 & -0.036 \\
    DNF                 & 0.010  & 0.700 & 0.690 & 0.520 & 0.349 & -0.171 \\
    f-DNF               & 0.010  & \textbf{0.710} & \textbf{0.700} & 0.520 & 0.503 & -0.017 \\
    \hline
	\end{tabular}}
\end{table}

It can be seen that DNF has a stronger capacity than VAE and NF to implement factor manipulation.
However, the DNF-based manipulation tends to cause larger distortion on other factors.
Factorial DNF has similar even better performance than DNF in terms of factor manipulation,
but causes very little distortion on other factors. Speech examples
can be found at http://project.cslt.org.

%
%
%
%
%
%

\section{Conclusions}
\label{sec:con}

This paper presented a speech information factorization method based on a novel deep generative model that we called
factorial discriminative normalization flow.
Qualitative and quantitative experimental results show that compared to all other models,
the proposed factorial DNF can retain the class structure corresponding to multiple information factors,
and changing one factor will cause little distortion on other factors.
This demonstrates that factorial DNF can well factorize speech signal into different information factors.
Future work will test factorial DNF on larger datasets, and establish general theories for deep generative
factorization.


\bibliographystyle{IEEEbib}
\bibliography{refs}

\end{document}